# Efficiently expressing feasibility problems in Linear Systems, as feasibility problems in Asymptotic-Linear-Programs

Deepak Ponvel Chermakani

deepakc@pmail.ntu.edu.sg  deepakc@e.ntu.edu.sg  deepakc@ed-alumni.net  deepakc@myfastmail.com  deepakc@usa.com

***Abstract: -*** **We present a polynomial-time algorithm that obtains a set of Asymptotic Linear Programs (ALPs) from a given linear system S, such that one of these ALPs admits a feasible solution if and only if S admits a feasible solution. We also show how to use the same algorithm to determine whether or not S admits a non-trivial solution for any desired subset of its variables. S is allowed to consist of linear constraints over real variables with integer coefficients, where each constraint has either a lesser-than-or-equal-to ($\leq$), or a lesser-than ($<$), or a not-equal-to ($\neq$) relational operator. Each constraint of the obtained ALPs has a lesser-than-or-equal-to ($\leq$) relational operator, and the coefficients of its variables vary linearly with respect to the time parameter that tends to positive infinity.**

## 1. Introduction

In our previous paper [1], we showed how to efficiently convert any given linear system with simultaneous constraints having lesser-than-or-equal-to ($\leq$) and lesser-than ($<$) relational operators, into an Asymptotic Linear Program (ALP), such that (The ALP has a feasible solution) $\leftrightarrow$ (The given linear system has a non-trivial feasible solution). In that paper [1], we showed that one way of modeling $n$ Inequations (i.e. constraints with not-equal-to relational operators), was to iteratively consider the rest of the constraints with two cases (for example, for $x\neq 0$, consider the remaining constraints separately with $x<0$ and $x>0$), which would lead to an overall exponential complexity of $O(2^n)$. We posed an open question, on efficiently (i.e. within polynomial-time) modeling Inequations as an ALP. In this paper, we show this is possible in $O(n^2)$ time.

## 2. The foundation for efficiently modeling Inequations

**Definition:** Let $<y_1, y_2, \ldots y_N>$, $<x_1, x_2, \ldots x_N>$ and $<z_1, z_2, \ldots z_N>$ be three vectors of real variables, and let $K$ be a real variable. Let the variables of these three vectors be related as follows:

$y_i$ = (Summation($x_j$, over all integers $j$ in $[1,N]$, and $j \neq i$)), for all integers $i$ in $[1,N]$.

$x_i = (z_i/(K+i))$, for all integers $i$ in $[1,N]$.

For example, for $N=5$:

$$y_1 = \phantom{x_1 +} x_2 + x_3 + x_4 + x_5$$
$$y_2 = x_1 \phantom{+ x_2} + x_3 + x_4 + x_5$$
$$y_3 = x_1 + x_2 \phantom{+ x_3} + x_4 + x_5$$
$$y_4 = x_1 + x_2 + x_3 \phantom{+ x_4} + x_5$$
$$y_5 = x_1 + x_2 + x_3 + x_4$$
$$x_1 = z_1/(K+1)$$
$$x_2 = z_2/(K+2)$$
$$x_3 = z_3/(K+3)$$
$$x_4 = z_4/(K+4)$$
$$x_5 = z_5/(K+5)$$

We will conveniently assume that $N>1$, because if there is only one not-equal-to constraint (say $t\neq 0$) in a linear system, one can easily solve the system by considering the remaining problem with two cases ($t<0$) and then with ($t>0$).

We now state and prove two Theorems, which will form the foundation for efficiently modeling inequations as an ALP.

**Theorem-1:** For all real values of the elements of vector $<y_1, y_2, \ldots y_N>$, the following statement is true: - (There exists a positive real $\gamma$, such that for all $K > \gamma$, there exists a real solution to the vectors $<x_1, x_2, \ldots x_N>$ and $<z_1, z_2, \ldots z_N>$)

**Proof:** For $K>0$, it is trivial to see that ($<x_1, x_2, \ldots x_N>$ has a real solution) $\leftrightarrow$ ($<z_1, z_2, \ldots z_N>$ has a real solution). Next, to prove that ($<x_1, x_2, \ldots x_N>$ has a real solution) for all real values of $<y_1, y_2, \ldots y_N>$, we need to show that the determinant of square matrix A (see Figure-1) defined by $\{a_{i,j}$ = $1$ if $i \neq j$,

$= 0$ if $i=j$, for all integers $i$ and $j$ in $[1,N]\}$, is non-zero.

|   |   |   |   |   |   |   |   |
|---|---|---|---|---|---|---|---|
| 0 | 1 | 1 | 1 | ... | ... | 1 | 1 |
| 1 | 0 | 1 | 1 | ... | ... | 1 | 1 |
| 1 | 1 | 0 | 1 | ... | ... | 1 | 1 |
| 1 | 1 | 1 | 0 | ... | ... | 1 | 1 |
| ... | ... | ... | ... | ... | ... | ... | ... |
| ... | ... | ... | ... | ... | ... | ... | ... |
| 1 | 1 | 1 | 1 | 1 | 1 | 0 | 1 |
| 1 | 1 | 1 | 1 | 1 | 1 | 1 | 0 |

**Figure-1:** The Determinant of square-Matrix A of dimension $N$

To show that the determinant of Matrix A is indeed non-zero, iteratively apply the following rule to Matrix A:
Row$_i$ = (Summation (Row$_j$, over all integers $j$ in $[i+1,N]$) $-$ $(N-i-1)$Row$_i$), for all integers $i$ in $[1,(N-2)]$. We then get the Matrix B defined by (for all integers $j$ in $[1,N]$):

$\{b_{i,j}$ (for all integers $i$ in $[1,(N-2)]$)$\}$      = $(N-i)$   if $i=j$,
     =1   if $i>j$,
     = 0   if $i<j$.$\}$, and,

$\{b_{i,j}$ (for all integers $i$ in $[(N-1),N]$)$\}$      =1   if $i \neq j$,
     = 0   if $i=j$.$\}$.

This Matrix B is a left diagonal matrix, whose determinant is equal to $-(N-1)!$, which is obviously non-zero for $N>1$.
**Hence Proved**

**Theorem-2:** There exists a positive real $\gamma$, such that for all $K > \gamma$, the following statement is true:
((Atleast two elements of vector $<z_1, z_2, \ldots z_N>$ are non-zero) $\leftrightarrow$ (All elements of vector $<y_1, y_2, \ldots y_N>$ are non-zero))
**Proof**: Consider the two boolean statements within the 'if-and-only-if' in the Theorem to be P and Q. It is well known that to prove (P $\leftrightarrow$ Q), it suffices to prove two statements: (Q $\rightarrow$ P) and ((not Q) $\rightarrow$ (not P)). It is easy to see that (All the elements of $<z_1, z_2, \ldots z_N>$ are zero) $\rightarrow$ (All the elements of $<y_1, y_2, \ldots y_N>$ are zero). Also, if exactly one of the elements of $<z_1, z_2, \ldots z_N>$ is non-zero, this would mean that exactly $(N-1)$ elements of $<y_1, y_2, \ldots y_N>$ are non-zero, because each variable $z_i$ appears in the defining equations of $\{y_j$, for all integers $j$ in $[1,N]$, where $j \neq i\}$. To put it explicitly, we have:
$y_i$ = (Summation($z_j/(K+j)$), over all integers $j$ between $1$ and $N$, and $j \neq i$)), for all integers $i$ in $[1,N]$.
Thus ((not Q) $\rightarrow$ (not P)). Next, from Theorem-1 of paper [1], and the above defining relationship, it follows that (P $\rightarrow$ Q) for all $K > \gamma$, where $\gamma$ is a positive real that is a function of the elements of $<y_1, y_2, \ldots y_N>$.
**Hence Proved**

## 3. Converting Linear Feasibility Problems into an ALP

The Linear System we consider is a set of simultaneous linear constraints over a vector of real variables $<x_1, x_2, \ldots x_N>$ (i.e. each variable is initially allowed to take the values of zero, positive Reals, or negative Reals). We shall refer to our Linear System as S$_{linear}$, having $P$ linear constraints with lesser-than-or-equal-to relational operators, $Q$ linear constraints with lesser-than relational operators, and $R$ linear constraints with not-equal-to relational operators:

$a_{1,1} x_1 + a_{1,2} x_2 + \ldots + a_{1,N} x_N \leq p_1$
$a_{2,1} x_1 + a_{2,2} x_2 + \ldots + a_{2,N} x_N \leq p_2$
...
$a_{P,1} x_1 + a_{P,2} x_2 + \ldots + a_{P,N} x_N \leq p_P$
$b_{1,1} x_1 + b_{1,2} x_2 + \ldots + b_{1,N} x_N < q_1$
$b_{2,1} x_1 + b_{2,2} x_2 + \ldots + b_{2,N} x_N < q_2$
...
$b_{Q,1} x_1 + b_{Q,2} x_2 + \ldots + b_{Q,N} x_N < q_Q$
$c_{1,1} x_1 + c_{1,2} x_2 + \ldots + c_{1,N} x_N \neq r_1$
$c_{2,1} x_1 + c_{2,2} x_2 + \ldots + c_{2,N} x_N \neq r_2$
...
$c_{R,1} x_1 + c_{R,2} x_2 + \ldots + c_{R,N} x_N \neq r_R$

In $S_{linear}$, for all integers $i$ in $[1,N]$, for all integers $j$ in $[1,P]$, for all integers $k$ in $[1,Q]$, for all integers $l$ in $[1,R]$: $x_i$ is a real variable, and the elements of $\{a_{j,i}, p_j, b_{k,i}, q_k, c_{l,i}, r_l\}$ belong to the set of integers. $S_{linear}$ is able to express most linear systems, except linear discrete systems (for example, if $x$ is constrained to integer values).

Let $P_{linear}$ be the problem of deciding whether or not $S_{linear}$ admits a feasible solution. Our Algorithm for $P_{linear}$ is as follows:

<u>Step-1</u>: Replace each constraint having a lesser-than relational operator, with 2 simultaneous constraints. For, example, $(x < a)$ can be replaced with a set of constraints of the form $(((a-x) \geq e)\ AND\ ((Ke) \geq 1))$, where $e$ is a real variable introduced, and where $K$ is the time parameter of our ALP (i.e. a real number that is assumed to tend to positive infinity).

<u>Step-2</u>: $S_{linear}$ now consists of constraints with only lesser-than-or-equal-to and not-equal-to relational operators. Divide $S_{linear}$ into two sets of constraints: - $S_{linear\_subset\_without\_inequations}$ (that has constraints with only the lesser-than-or-equal-to operators) and $S_{linear\_subset\_inequations}$ (that has the $R$ constraints with only the not-equal-to operators).

<u>Step-3</u>: Write out the $R$ inequations as follows:
$$c_{1,1} x_1 + c_{1,2} x_2 + \ldots + c_{1,N} x_N = f_1$$
$$c_{2,1} x_1 + c_{2,2} x_2 + \ldots + c_{2,N} x_N = f_2$$
$$\ldots$$
$$c_{R,1} x_1 + c_{R,2} x_2 + \ldots + c_{R,N} x_N = f_R$$
$$f_1 \neq 0$$
$$f_2 \neq 0$$
$$\ldots$$
$$f_R \neq 0$$

<u>Step-4</u>: Write $f_i =$ (Summation($y_j$, over all integers $j$ between $1$ and $R$, and $j \neq i$)), for all integers $i$ in $[1,R]$. Also, write $(K+i)y_i = z_i$ for all integers $i$ in $[1,R]$. Here $<y_1, y_2, \ldots y_R>$ and $<z_1, z_2, \ldots z_R>$ are the vectors of real variables introduced.

<u>Step-5</u>: Write each of the constraints with equal-to operators obtained in Step-3 and Step-4 (and any other such constraints initially present in $S_{linear}$), as two simultaneous constraints with lesser-than-or-equal-to operators. For example, $(x = a)$ can be expressed as a set of two constraints $(((x-a) \leq 0)\ AND\ ((a-x) \leq 0))$. Add these constraints to $S_{linear\_subset\_without\_inequations}$.

<u>Step-6</u>: Consider $^RC_2$ cases ($= R(R-1)/2$ cases) by taking all possible combinations of $2$ elements from the vector $<z_1, z_2, \ldots z_R>$ to be not-equal-to-zero. For each of these $R(R-1)/2$ cases, there will be $4$ separate cases, involving each of these $2$ elements being $> 0$ and $< 0$. For example, if $z_2$ and $z_5$ are selected, we have $4$ separate cases: - $(-z_2<0, -z_5<0)$, $(-z_2<0, z_5<0)$, $(z_2<0, -z_5<0)$ and $(z_2<0, z_5<0)$. We thus have a total of $2R(R-1)$ separate cases.

<u>Step-7</u>: For each of these $2R(R-1)$ separate cases, convert the 2 constraints with lesser-than operators into constraints with lesser-than-or-equal-to operators using the procedure described in Step-1.

<u>Step-8</u>: Write $ALP_i$ as the union of $S_{linear\_subset\_without\_inequations}$ with the constraints with lesser-than-or-equal-to operators of $Case_i$ described above in Step-7, for all integers $i$ in $[1, (2R(R-1))]$.

<u>Step-9</u>: (For at least one of the integers $i$ in $[1, (2R(R-1))]$, $ALP_i$ is feasible) $\leftrightarrow$ ($S_{linear}$ is feasible).

A Note on the Aymptotic Linear Program (ALP)
An ALP [2][3][4] is a linear program, where the coefficients of the variables in the constraints are rational Polynomials involving a single real variable called the time parameter. The author of [4] proved that as this time parameter grows beyond a certain positive value, the Linear Program gets constant (i.e. steady-state) properties of feasibility or infeasibility. In other words, as this time parameter tends to positive infinity, the Asymptotic Linear Program becomes either feasible or infeasible.

Start of Example demonstrating Algorithm for $P_{linear}$:
Consider $S_{linear}$ to be defined by the following set of $7$ linear constraints over the real variable vector $<x_1, x_2, x_3>$:
$$a_{1,1} x_1 + a_{1,2} x_2 + a_{1,3} x_3 \leq p_1$$
$$a_{2,1} x_1 + a_{2,2} x_2 + a_{2,3} x_3 \leq p_2$$
$$b_{1,1} x_1 + b_{1,2} x_2 + b_{1,3} x_3 < q_1$$
$$b_{2,1} x_1 + b_{2,2} x_2 + b_{2,3} x_3 < q_2$$
$$c_{1,1} x_1 + c_{1,2} x_2 + c_{1,3} x_3 \neq r_1$$
$$c_{2,1} x_1 + c_{2,2} x_2 + c_{2,3} x_3 \neq r_2$$
$$c_{3,1} x_1 + c_{3,2} x_2 + c_{3,3} x_3 \neq r_3.$$

After Step-1, $S_{linear}$ becomes the following:
$$a_{1,1} x_1 + a_{1,2} x_2 + a_{1,3} x_3 \leq p_1$$
$$a_{2,1} x_1 + a_{2,2} x_2 + a_{2,3} x_3 \leq p_2$$
$$b_{1,1} x_1 + b_{1,2} x_2 + b_{1,3} x_3 - q_1 \leq -e$$
$$b_{2,1} x_1 + b_{2,2} x_2 + b_{2,3} x_3 - q_2 \leq -e$$
$$c_{1,1} x_1 + c_{1,2} x_2 + c_{1,3} x_3 \neq r_1$$
$$c_{2,1} x_1 + c_{2,2} x_2 + c_{2,3} x_3 \neq r_2$$

$c_{3,1} x_1 + c_{3,2} x_2 + c_{3,3} x_3 \neq r_3$
$Ke \geq 1$, where $e$ is the real variable introduced, and $K$ is the large positive real.

As per Step-2, $S_{\text{linear\_subset\_without\_inequations}}$ consists of the following:
$a_{1,1} x_1 + a_{1,2} x_2 + a_{1,3} x_3 \leq p_1$
$a_{2,1} x_1 + a_{2,2} x_2 + a_{2,3} x_3 \leq p_2$
$b_{1,1} x_1 + b_{1,2} x_2 + b_{1,3} x_3 - q_1 \leq -e$
$b_{2,1} x_1 + b_{2,2} x_2 + b_{2,3} x_3 - q_2 \leq -e$
$Ke \geq 1$,

and $S_{\text{linear\_subset\_inequations}}$ consists of the following:
$c_{1,1} x_1 + c_{1,2} x_2 + c_{1,3} x_3 - r_1 \neq 0$
$c_{2,1} x_1 + c_{2,2} x_2 + c_{2,3} x_3 - r_2 \neq 0$
$c_{3,1} x_1 + c_{3,2} x_2 + c_{3,3} x_3 - r_3 \neq 0$.

As per Step-3, we have:
$c_{1,1} x_1 + c_{1,2} x_2 + c_{1,3} x_3 - r_1 = f_1$
$c_{2,1} x_1 + c_{2,2} x_2 + c_{2,3} x_3 - r_2 = f_2$
$c_{3,1} x_1 + c_{3,2} x_2 + c_{3,3} x_3 - r_3 = f_3$
$f_1 \neq 0$
$f_2 \neq 0$
$f_3 \neq 0$.

As per Step-4, we have:
$f_1 = y_2 + y_3$
$f_2 = y_1 + y_3$
$f_3 = y_1 + y_2$
$(K+1)y_1 = z_1$
$(K+2)y_2 = z_2$
$(K+3)y_3 = z_3$, where $<y_1, y_2, y_3>$ and $<z_1, z_2, z_3>$ are the vectors of real variables introduced.

As per Step-5, $S_{\text{linear\_subset\_without\_inequations}}$ now becomes:
$a_{1,1} x_1 + a_{1,2} x_2 + a_{1,3} x_3 \leq p_1$
$a_{2,1} x_1 + a_{2,2} x_2 + a_{2,3} x_3 \leq p_2$
$b_{1,1} x_1 + b_{1,2} x_2 + b_{1,3} x_3 - q_1 \leq -e$
$b_{2,1} x_1 + b_{2,2} x_2 + b_{2,3} x_3 - q_2 \leq -e$
$1 - Ke \leq 0$
$c_{1,1} x_1 + c_{1,2} x_2 + c_{1,3} x_3 - r_1 - f_1 \leq 0$
$c_{2,1} x_1 + c_{2,2} x_2 + c_{2,3} x_3 - r_2 - f_2 \leq 0$
$c_{3,1} x_1 + c_{3,2} x_2 + c_{3,3} x_3 - r_3 - f_3 \leq 0$
$-c_{1,1} x_1 - c_{1,2} x_2 - c_{1,3} x_3 + r_1 + f_1 \leq 0$
$-c_{2,1} x_1 - c_{2,2} x_2 - c_{2,3} x_3 + r_2 + f_2 \leq 0$
$-c_{3,1} x_1 - c_{3,2} x_2 - c_{3,3} x_3 + r_3 + f_3 \leq 0$
$y_2 + y_3 - f_1 \leq 0$
$y_1 + y_3 - f_2 \leq 0$
$y_1 + y_2 - f_3 \leq 0$
$-y_2 - y_3 + f_1 \leq 0$
$-y_1 - y_3 + f_2 \leq 0$
$-y_1 - y_2 + f_3 \leq 0$
$(K+1)y_1 - z_1 \leq 0$
$(K+2)y_2 - z_2 \leq 0$
$(K+3)y_3 - z_3 \leq 0$
$-(K+1)y_1 + z_1 \leq 0$
$-(K+2)y_2 + z_2 \leq 0$
$-(K+3)y_3 + z_3 \leq 0$

As per Step-6, we have a total of 12 cases:
Case$_1$: $z_1 < 0$, $z_2 < 0$     Case$_2$: $z_1 < 0$, $z_3 < 0$
Case$_3$: $z_2 < 0$, $z_3 < 0$     Case$_4$: $z_1 < 0$, $z_2 > 0$
Case$_5$: $z_1 < 0$, $z_3 > 0$     Case$_6$: $z_2 < 0$, $z_3 > 0$
Case$_7$: $z_1 > 0$, $z_2 < 0$     Case$_8$: $z_1 > 0$, $z_3 < 0$
Case$_9$: $z_2 > 0$, $z_3 < 0$     Case$_{10}$: $z_1 > 0$, $z_2 > 0$

Case$_{11}$: $z_1 > 0$, $z_3 > 0$     Case$_{12}$: $z_2 > 0$, $z_3 > 0$

As per Step-7, the constraints with lesser-than operators can be converted to constraints with lesser-than-or-equal-to operators. So we have:

Case$_1$: $z_1 \leq e$, $z_2 \leq e$, $1 \leq Ke$        Case$_2$: $z_1 \leq e$, $z_3 \leq e$, $1 \leq Ke$
Case$_3$: $z_2 \leq e$, $z_3 \leq e$, $1 \leq Ke$        Case$_4$: $z_1 \leq e$, $z_2 \geq e$, $1 \leq Ke$
Case$_5$: $z_1 \leq e$, $z_3 \geq e$, $1 \leq Ke$        Case$_6$: $z_2 \leq e$, $z_3 \geq e$, $1 \leq Ke$
Case$_7$: $z_1 \geq e$, $z_2 \leq e$, $1 \leq Ke$        Case$_8$: $z_1 \geq e$, $z_3 \leq e$, $1 \leq Ke$
Case$_9$: $z_2 \geq e$, $z_3 \leq e$, $1 \leq Ke$        Case$_{10}$: $z_1 \geq e$, $z_2 \geq e$, $1 \leq Ke$
Case$_{11}$: $z_1 \geq e$, $z_3 \geq e$, $1 \leq Ke$      Case$_{12}$: $z_2 \geq e$, $z_3 \geq e$, $1 \leq Ke$

As per Step-8, we generate a set of 12 separate ALPs. {ALP$_1$, ALP$_2$, … ALP$_{11}$, ALP$_{12}$}. Here ALP$_i$ is the union of S$_{linear\_subset\_without\_inequations}$ with the constraints of Case$_i$ described above in Step-7, for all integers $i$ in [1, 12]. For example, ALP$_{11}$ is shown below:

$a_{1,1} x_1 + a_{1,2} x_2 + a_{1,3} x_3 \leq p_1$
$a_{2,1} x_1 + a_{2,2} x_2 + a_{2,3} x_3 \leq p_2$
$b_{1,1} x_1 + b_{1,2} x_2 + b_{1,3} x_3 - q_1 \leq -e$
$b_{2,1} x_1 + b_{2,2} x_2 + b_{2,3} x_3 - q_2 \leq -e$
$c_{1,1} x_1 + c_{1,2} x_2 + c_{1,3} x_3 - r_1 - f_1 \leq 0$
$c_{2,1} x_1 + c_{2,2} x_2 + c_{2,3} x_3 - r_2 - f_2 \leq 0$
$c_{3,1} x_1 + c_{3,2} x_2 + c_{3,3} x_3 - r_3 - f_3 \leq 0$
$-c_{1,1} x_1 - c_{1,2} x_2 - c_{1,3} x_3 + r_1 + f_1 \leq 0$
$-c_{2,1} x_1 - c_{2,2} x_2 - c_{2,3} x_3 + r_2 + f_2 \leq 0$
$-c_{3,1} x_1 - c_{3,2} x_2 - c_{3,3} x_3 + r_3 + f_3 \leq 0$
$y_2 + y_3 - f_1 \leq 0$
$y_1 + y_3 - f_2 \leq 0$
$y_1 + y_2 - f_3 \leq 0$
$-y_2 - y_3 + f_1 \leq 0$
$-y_1 - y_3 + f_2 \leq 0$
$-y_1 - y_2 + f_3 \leq 0$
$(K+1)y_1 - z_1 \leq 0$
$(K+2)y_2 - z_2 \leq 0$
$(K+3)y_3 - z_3 \leq 0$
$-(K+1)y_1 + z_1 \leq 0$
$-(K+2)y_2 + z_2 \leq 0$
$-(K+3)y_3 + z_3 \leq 0$
$e - z_1 \leq 0$
$e - z_3 \leq 0$
$1 - Ke \leq 0$

Finally, in Step-9, we determine feasibility of the 12 ALPs. (Atleast one of {ALP$_1$, ALP$_2$, … ALP$_{11}$, ALP$_{12}$} is feasible) ↔ (S$_{linear}$ is feasible).

End of Example demonstrating Algorithm for P$_{linear}$.

## 4. Deciding non-triviality of a subset of variables of S$_{linear}$

(A vector of reals <$\mu_1$, $\mu_2$, … $\mu_N$> is said to be non-trivial) ↔ (For at least one integer $i$ in [1,N], $\mu_i \neq 0$). If it is desired to determine whether or not S$_{linear}$ permits a non-trivial solution for a subset of the variables of the vector <$x_1$, $x_2$, … $x_N$>, we can introduce an additional constraint with a not-equal-to operator using Theorem-1 of the paper [1]. For example, if it is desired to determine whether or not S$_{linear}$ permits a non-trivial solution for <$x_2$, $x_5$, $x_{13}$, $x_N$>, we have:
$(x_2 / (K+1)) + (x_5 / (K+2)) + (x_{13} / (K+3)) + (x_N / (K+4))) \neq 0$, which may be expressed using the following set of 5 simultaneous constraints, where <$w_2$, $w_5$, $w_{13}$, $w_N$> is the vector of variables introduced:

$w_2 + w_5 + w_{13} + w_N \neq 0$
$x_2 = (K+1) w_2$
$x_5 = (K+2) w_5$
$x_{13} = (K+3) w_{13}$
$x_N = (K+N) w_N$

We call the union of S$_{linear}$ and the above set of 5 constraints, as S$_{linear\_non\_trivial\_2\_5\_13\_N}$, which is shown next:

$$a_{1,1} x_1 + a_{1,2} x_2 + \ldots + a_{1,N} x_N \leq p_1$$
$$a_{2,1} x_1 + a_{2,2} x_2 + \ldots + a_{2,N} x_N \leq p_2$$
$$\ldots$$
$$a_{P,1} x_1 + a_{P,2} x_2 + \ldots + a_{P,N} x_N \leq p_P$$
$$b_{1,1} x_1 + b_{1,2} x_2 + \ldots + b_{1,N} x_N < q_1$$
$$b_{2,1} x_1 + b_{2,2} x_2 + \ldots + b_{2,N} x_N < q_2$$
$$\ldots$$
$$b_{Q,1} x_1 + b_{Q,2} x_2 + \ldots + b_{Q,N} x_N < q_Q$$
$$c_{1,1} x_1 + c_{1,2} x_2 + \ldots + c_{1,N} x_N \neq r_1$$
$$c_{2,1} x_1 + c_{2,2} x_2 + \ldots + c_{2,N} x_N \neq r_2$$
$$\ldots$$
$$c_{R,1} x_1 + c_{R,2} x_2 + \ldots + c_{R,N} x_N \neq r_R$$
$$w_2 + w_5 + w_{13} + w_N \neq 0$$
$$x_2 = (K+1) w_2$$
$$x_5 = (K+2) w_5$$
$$x_{13} = (K+3) w_{13}$$
$$x_N = (K+N) w_N$$

We now apply the algorithm for $P_{linear}$ to $S_{linear\_non\_trivial\_2\_5\_13\_N}$. ($S_{linear\_non\_trivial\_2\_5\_13\_N}$ has a feasible solution) ↔ ($S_{linear}$ permits a non-trivial feasible solution for $<x_2, x_5, x_{13}, x_N>$).

## 5. Conclusion

In this paper, we developed the foundations for expressing the feasibility of a set of Inequations within Linear Systems, using ALPs, within polynomial time, thus answering an open question we posed in our previous paper. We then developed a polynomial-time algorithm to express as ALP problems, the feasibility of a set of linear constraints over real variables with integer coefficients, each constraint having one of 4 types of relational operators (=, ≤, < and ≠). The resulting ALP problems have linear constraints (with the ≤ operator) over real variables with coefficients that vary linearly with the time parameter $K$ that tends to positive infinity. We also showed how to efficiently (within polynomial-time) convert the question of whether or not the linear system allows a subset of its variables to be non-trivial, into the question of whether or not another linear system (with =, ≤, < and ≠ relational operators) has a feasible solution, thus allowing our polynomial-time algorithm to be used for determining feasibility of the non-trivial solution of the desired subset of variables of the original linear system.

## 6. Future Work

If it possible to express (within polynomial-time) the question of whether or not linear constraints over binary-variables (i.e. the variables are allowed to take the values of either 0 or 1), as ALPs, this would prove that Aymptotic-Linear-Programming is NP-hard. So this is an important open problem. Another open problem is whether or not a weakly-polynomial-time algorithm exists for Aymptotic-Linear-Programming (just as weakly-polynomial-time algorithms already exist for Ordinary-Linear-Programming).

**About the Author**
I, Deepak Ponvel Chermakani, wrote this paper out of my own interest and initiative, during my spare time. In Sep-2010, I completed a fulltime one year Master Degree in *Operations Research with Computational Optimization* from University of Edinburgh UK (www.ed.ac.uk). In Jul-2003, I completed a fulltime four year Bachelor Degree in *Electrical and Electronic Engineering*, from Nanyang Technological University Singapore (www.ntu.edu.sg). In Jul-1999, I completed fulltime high schooling from National Public School in Bangalore in India.